\documentclass[a4paper,11pt]{article}
\usepackage{pos}
\usepackage{subcaption}

\newcommand{\ahat}{\ensuremath{\widehat a} }
\newcommand{\ahatdag}{\ensuremath{\widehat a^{\dag}} }
\newcommand{\cN}{\ensuremath{\mathcal N} }
\newcommand{\be}{\ensuremath{\beta} }
\newcommand{\chidag}{\ensuremath{\chi^{\dag}} }
\newcommand{\La}{\ensuremath{\Lambda} }
\newcommand{\la}{\ensuremath{\lambda} }
\newcommand{\Om}{\ensuremath{\Omega} }
\newcommand{\ket}[1]{\ensuremath{\left| #1 \right\rangle} }
\newcommand{\bra}[1]{\ensuremath{\left\langle #1 \right|} }
\newcommand{\EV}[2]{\ensuremath{\langle #1 \left| #2 \right| #1 \rangle} }
\newcommand{\eq}[1]{Eq.~\ref{#1}}
\newcommand{\fig}[1]{Fig.~\ref{#1}}
\newcommand{\secref}[1]{Section~\ref{#1}}
\newcommand{\refcite}[1]{Ref.~\cite{#1}}

\title{Exploring lattice supersymmetry \\ \hfill with variational quantum deflation}
\ShortTitle{Exploring lattice supersymmetry with variational quantum deflation}

\author*{David Schaich}
\author{Christopher Culver$^{\dag}$}
\affiliation{Department of Mathematical Sciences, University of Liverpool, Liverpool L69 7ZL, United Kingdom} 
\emailAdd{david.schaich@liverpool.ac.uk}

\FullConference{40th International Symposium on Lattice Field Theory \\ 31 July -- 4 August 2023 \\ Fermi National Accelerator Laboratory}

\abstract{ 
  Lattice studies of spontaneous supersymmetry breaking suffer from a sign problem that in principle can be evaded through novel methods enabled by quantum computing.
  Focusing on lower-dimensional lattice systems with more modest resource requirements, in particular the $\cN = 1$ Wess--Zumino model in 1+1 dimensions, we are exploring ways quantum computing could be used to study spontaneous supersymmetry breaking.
  A particularly promising recent development is to apply the variational quantum deflation algorithm, which generalizes the variational quantum eigensolver so as to resolve multiple low-energy states.
}

\begin{document}
\maketitle

\section{\label{sec:intro}Introduction} 
Efforts to use lattice field theory to non-perturbatively analyze supersymmetric systems have a long history, motivated by the many prominent roles supersymmetry plays in modern theoretical physics.
These include famous proposed extensions of the standard model, and `holographic' dualities with higher-dimensional theories of quantum gravity.
In addition, supersymmetry is a valuable tool to improve our understanding of quantum field theory more generally, and has been used to obtain insight into non-perturbative phenomena including confinement, conformality, and dynamical symmetry breaking.

While it is famously difficult for lattice field theory calculations in discrete space-time to preserve the supersymmetric extension of the Poincar\'e algebra, the situation becomes simpler for lower-dimensional systems --- see Refs.~\cite{Kadoh:2016eju, Bergner:2016sbv, Schaich:2022xgy} for recent reviews.
In particular, considering the hamiltonian formulation in which only space is discretized while time remains continuous, it is possible to construct an exactly supersymmetric discretization of the $\cN = 1$ Wess--Zumino model in 1+1~dimensions, as we will review below.
Even for this supersymmetric lattice system, however, Monte Carlo importance sampling studies are often obstructed by severe sign problems.
These arise, for example, if we wish to analyze the real-time evolution of the system, or numerically investigate dynamical supersymmetry breaking~\cite{Bergner:2016sbv, Schaich:2022xgy}. 

We have recently begun exploring ways in which quantum computing might provide a novel means to study such phenomena, without suffering from sign problems~\cite{Culver:2021rxo, Culver:2023iif}.
This research remains highly exploratory at present, given the modest numbers of qubits and relatively high error rates characterizing existing and near-future quantum devices.
The current status of quantum technologies is widely described as the Noisy Intermediate-Scale Quantum (NISQ) era~\cite{Preskill:2018nisq}, in contrast to future fault-tolerant computing that requires improved resources in order to enable quantum error correction.

Even in the NISQ era, it is important to begin investigating how quantum computing could contribute to lattice field theory research, and testing potential reformulations and algorithms that may be needed for this approach to be successful~\cite{Alexeev:2020xrq}.
These initial studies are limited to small systems and shallow circuit depths, allowing them to be checked by comparison with classical diagonalization.
The focus of this proceedings is to explore the use of `hybrid' variational algorithms, specifically the Variational Quantum Eigensolver (VQE)~\cite{McClean:2016vqe} and Variational Quantum Deflation (VQD)~\cite{Higgott:2019vqd}.
This approach employs a quantum circuit to efficiently evaluate an `objective function' at each step of an iterative classical optimization routine, providing a more modest quantum computation that is better suited for existing and near-future hardware.

In this spirit, we are currently investigating dynamical supersymmetry breaking in the (1+1)-dimensional $\cN = 1$ Wess--Zumino model, which is arguably the simplest supersymmetric quantum field theory.
This system has been the subject of lattice investigations from a variety of approaches~\cite{Kadoh:2016eju}, ranging from the traditional lagrangian formulation~\cite{Catterall:2003ae, Wozar:2011gu} to the continuous-time hamiltonian formulation~\cite{Beccaria:2001qm, Beccaria:2003gt, Beccaria:2004pa, Beccaria:2004ds}, the fermion loop formulation~\cite{Steinhauer:2014yaa}, and tensor network formulations~\cite{Kadoh:2018hqq, Meurice:2020pxc}.
We begin in the next section by briefly summarizing the hamiltonian formulation of this lattice theory, which possesses an exact supersymmetry even at non-zero lattice spacing.
In \secref{sec:results} we then discuss the latest results from our VQE and VQD analyses of dynamical supersymmetry breaking in the Wess--Zumino model, before concluding in \secref{sec:conc} with a discussion of our planned next steps.

\section{Lattice Wess--Zumino model} 
\subsection{\label{sec:lattice}Hamiltonian lattice formulation} 
The $\cN = 1$ Wess--Zumino model in 1+1 dimensions adds a two-component fermion $\psi$ to simple scalar $\phi^4$ theory.
Imposing supersymmetry ensures that the scalar and the fermion have the same mass, and also relates the couplings in the Yukawa and $\phi^4$ interactions.
This can be seen directly from the lattice hamiltonian~\cite{Beccaria:2001qm, Beccaria:2004pa}
\begin{equation}
  \label{eq:ham}
  \begin{split}
    H = \sum_{n=1}^{N_s} \Bigg[\frac{p_n^2}{2a} + \frac{a}{2}\left(\frac{\phi_{n+1} - \phi_{n-1}}{2a}\right)^2 & + \frac{a}{2}[V(\phi_n)]^2 + aV(\phi_n)\frac{\phi_{n+1} - \phi_{n-1}}{2a} \\
                                                                                                               & + (-1)^n V'(\phi_n)\left(\chidag_n\chi_n - \frac{1}{2}\right) + \frac{1}{2a}\left(\chidag_n\chi_{n+1} + \chidag_{n+1}\chi_n\right) \Bigg],
  \end{split}
\end{equation}
which is obtained by replacing the two fermion components $\psi_{1,n}$ and $\psi_{2,n}$ with creation and annihilation operators $\chidag_n$ and $\chi_n$ defined by
\begin{align*}
  \psi_{1,n} & = \frac{1 - i(-1)^n}{2i^n}\left(\chidag_n + i\chi_n\right) &
  \psi_{2,n} & = \frac{1 + i(-1)^n}{2i^n}\left(\chidag_n - i\chi_n\right).
\end{align*}
In these expressions, $n$ labels each lattice site, `$a$' is the lattice spacing, $p_n$ is the momentum conjugate to $\phi_n$, and $V(\phi_n)$ is a real `prepotential'.
Different choices for the prepotential produce Wess--Zumino systems that may exhibit qualitatively different behavior, including either the presence or absence of dynamical supersymmetry breaking.
Note that only the space dimension is discretized into a total of $N_s$ sites, while time remains continuous. 
We will impose either Dirichlet boundary conditions (BCs) or periodic BCs in our work presented below.

From \eq{eq:ham} we can see that $m\phi_n \subset V(\phi_n)$ would provide mass terms for both the boson and fermion, while $\sqrt{\la} \phi_n^2 \subset V(\phi_n)$ would lead to both $\phi^4$ and Yukawa interactions.
While this is consistent with supersymmetry, it is more subtle to establish that this lattice system remains exactly supersymmetric at non-zero lattice spacing.
This is elegantly done in Refs.~\cite{Beccaria:2001qm, Beccaria:2004pa} by showing that the lattice hamiltonian above can be expressed as the square of the discretized supercharge
\begin{equation*}
  Q = \frac{1}{\sqrt{a}}\sum_{n=1}^{N_s} \left[p_n \psi_{1,n}-\left(\frac{\phi_{n+1} - \phi_{n-1}}{2}+aV(\phi_n)\right)\psi_{2,n}\right].
\end{equation*}
That is, $H = Q^2$, as required by the super-Poincar\'e algebra.
This relation has several important consequences that we can exploit in our study of the Wess--Zumino model.
Specifically, all states in the system must have non-negative energy, $E_{\Psi} = \EV{\Psi}{H} = |Q\ket{\Psi}\!|^2 \geq 0$.
In addition, all energy eigenstates with a strictly positive $E_{\Psi} > 0$ come in pairs; only a supersymmetric ground state with $Q\ket{\Om} = 0 \to E_0 = 0$ can appear without a degenerate partner.
Finally, supersymmetry is spontaneously broken if and only if the minimum-energy ground state has a non-zero energy $E_0 > 0$.
This informs how we use the VQE and VQD to investigate spontaneous supersymmetry breaking, as we will discuss in more detail in the next section.

Before turning to those variational quantum algorithms, let's specify the particular prepotentials that we will consider in this proceedings.
The simplest non-trivial case is the linear prepotential
\begin{equation}
  \label{eq:linear}
  V(\phi_n) = \phi_n,
\end{equation}
which produces a free (non-interacting) supersymmetric field theory in which the boson and fermion both have a mass that we set to $m = 1$ for simplicity.
Not surprisingly, supersymmetry should be preserved for this free theory.

A more interesting case is the family of quadratic prepotentials
\begin{equation}
  \label{eq:quad}
  V(\phi_n) = c + \phi_n^2,
\end{equation}
with free parameter $c \in \mathbb R$, which was studied in Refs.~\cite{Beccaria:2001qm, Beccaria:2004pa, Beccaria:2004ds}.
This introduces $\phi^4$ and Yukawa interactions as discussed above, for which we again set the coupling $\la = 1$ for simplicity.
For sufficiently large $c > c_0$, this prepotential is expected to result in dynamical supersymmetry breaking.
In Refs.~\cite{Beccaria:2004pa, Beccaria:2004ds}, this critical value was found to be $c_0 \approx -0.5$.
In general, for polynomial prepotentials such as Eqs.~\ref{eq:linear}--\ref{eq:quad}, tree-level analyses suggest that supersymmetry should remain preserved when the highest power of $\phi_n$ in the polynomial (i.e., its order $q$) is odd, but may break spontaneously when $q$ is even~\cite{Beccaria:2004pa}. 

\subsection{\label{sec:qubit}Qubitization for quantum computing} 
In addition to the discretization of the space dimension leading to \eq{eq:ham}, in order to simulate this $\cN = 1$ lattice Wess--Zumino model using a quantum device we also need to map the bosonic and fermionic degrees of freedom to qubits.
This is straightforward for fermions in 1+1 dimensions, where we can use the usual Jordan--Wigner transformation,
\begin{align*}
  \chidag_n & = \frac{X_n - iY_n}{2} \qquad &
     \chi_n & = \frac{X_n + iY_n}{2}.
\end{align*}
Here $X_n$ and $Y_n$ represent the respective Pauli gate acting on the qubit corresponding to the fermion on the $n$th lattice site, $\psi_n$.

It is more complicated to map the scalar $\phi_n$ to two-state qubits.
(This may motivate future work considering continuous-variable quantum computing for the Wess--Zumino model~\cite{Jha:2023ecu}.)
At each lattice site, there is an infinite-dimensional Hilbert space that we need to truncate in order to work with a finite number of qubits.
To carry out this truncation, we represent each bosonic degree of freedom $\phi_n$ in the harmonic oscillator basis and impose a hard cutoff \La on the number of modes retained at each site.
This introduces an explicit breaking of the lattice supersymmetry discussed above, which is only recovered when the cutoff is removed, $\La \to \infty$.
By comparing results for different truncations $\La$, we will be able to assess the severity of the explicit supersymmetry breaking and disentangle it from the dynamical breaking of interest.

In this setup, we can write the bosonic raising and lowering operators as
\begin{align*}
     \ahat_n & = \sum_{\ell = 0}^{\La - 2} \sqrt{\ell + 1}\ket{\ell}\!\bra{\ell + 1} \qquad &
  \ahatdag_n & = \sum_{\ell = 0}^{\La - 2} \sqrt{\ell + 1}\ket{\ell + 1}\!\bra{\ell},
\end{align*}
and express each basis state as a tensor product of $N_b$ qubit states,
\begin{equation*}
  \ket{j} = \ket{b_0} \ket{b_1} \cdots \ket{b_{N_b - 1}}.
\end{equation*}
Here the integer label $j = \sum_{i = 0}^{N_b - 1} b_i 2^i$ is decomposed into $N_b$ binary factors associated with the $N_b$ qubits assigned to this particular $\phi_n$ on the $n$th lattice site.
In this proceedings we consider only $\La = 2^{N_b}$ such that $N_b = \log_2 \La$, though this can be generalized to $N_b = \lceil \log_2 \La \rceil$ in order to work with cutoffs that are not exact powers of two~\cite{Culver:2021rxo, Culver:2023iif}.
With this binary encoding, all bosonic operators
\begin{equation*}
  \ket{n}\!\bra{n'} = \otimes_{i = 0}^{N_b - 1} \ket{b_i} \!\bra{b_i'}
\end{equation*}
decompose into so-called Pauli strings via the relations
\begin{align*}
  \ket{0}\!\bra{1} & = \frac{X + iY}{2} \qquad &
  \ket{1}\!\bra{0} & = \frac{X - iY}{2} \\
  \ket{0}\!\bra{0} & = \frac{1 + Z}{2} \qquad &
  \ket{1}\!\bra{1} & = \frac{1 - Z}{2}.
\end{align*}
These Pauli strings correspond to the quantum gates that act on the qubits in order to implement the operator.
(Additional quantum gates are needed to implement the variational wavefunction ansatz, as discussed below.)
In total we need $N_q \equiv N_s \times (N_b + 1)$ qubits to analyze the $\cN = 1$ Wess--Zumino model on a lattice with $N_s$ sites with bosonic cutoff $\La = 2^{N_b}$.

\section{\label{sec:results}Results from variational quantum analyses} 
\subsection{Variational quantum eigensolver} 
As discussed already in \secref{sec:intro}, the small systems that we consider here are within the reach of classical diagonalization.
This makes it possible for us to use classical calculations to validate the results from quantum computations, explore the limitations of the latter and investigate how they may be improved by various quantum error mitigation strategies.
In addition, in order to accelerate our exploratory code development and testing at this stage of our work, the variational quantum results presented here are all obtained from classical simulations of quantum devices, using IBM's open-source Qiskit software development kit~\cite{Qiskit}.
Our Qiskit-based code for the $\cN = 1$ Wess--Zumino model is available at \href{https://github.com/daschaich/WessZumino}{github.com/daschaich/WessZumino}.

Also in \secref{sec:intro} we briefly introduced the VQE and VQD algorithms~\cite{McClean:2016vqe, Higgott:2019vqd}.
The VQE is quite well known by now, in particular as a way to approximate the ground state of a quantum system by using the energy as the objective function that it minimizes.
From the discussion in \secref{sec:lattice}, we can now appreciate how this enables us to investigate dynamical supersymmetry breaking in the Wess--Zumino model: If supersymmetry is unbroken the ground state must have $E_0 = 0$, and any non-zero ground-state energy implies spontaneously broken supersymmetry.
As a reminder, this is complicated by the finite cutoff \La that introduces explicit supersymmetry breaking, requiring us to consider several different cutoffs to disentangle these two effects.

We will show below that this disentangling can be done, but first we need to discuss the wavefunction ansatz $\ket{\Psi(\theta_i)}$ used by both the VQE and VQD.
This wavefunction depends on the $N_P$ parameters $\theta_i$, which are iteratively adjusted by the classical optimization routine so as to minimize the objective function (the energy).
We have tested several optimizers provided by Qiskit, among which the Constrained Optimization BY Linear Approximation (COBYLA) method demonstrated the best performance in our tests.
All results shown here use COBYLA.

\begin{figure}
  \centering
  \includegraphics[width=0.7\linewidth]{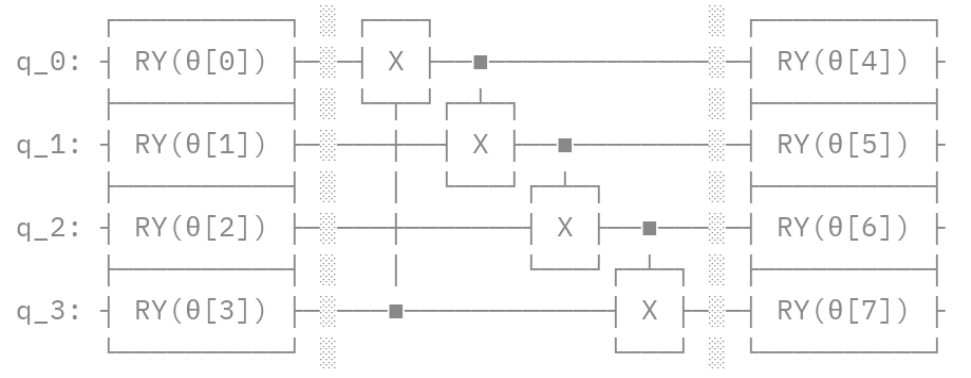}
  \caption{\label{fig:ansatz}A single repetition of the `circular' \texttt{RealAmplitudes} circuit we use to implement our variational wavefunction ansatz $\ket{\Psi(\theta_i)}$.  The four qubits here could represent either $N_s = 2$ sites with $\La = 2$ or a single site with $\La = 8$.}
\end{figure}

As the number of variational parameters increases, the optimization becomes more computationally difficult, requiring that we balance the expressivity of our ansatz against the computational cost.
For a small system (with $N_s = 2$ sites, $\La = 16$ and Dirichlet BCs), we tested various ans{\"a}tze based on \texttt{RealAmplitudes} circuits provided by Qiskit, which alternate CNOT gates and parameterized $Y$ rotations that depend on the variational $\theta_i$.
We obtained the best performance upon using $N_{\text{rep}} = 2$ repetitions of the `circular' \texttt{RealAmplitudes} circuit, one repetition of which is shown in \fig{fig:ansatz} for illustration.
Because only a single layer of rotations is done in between each repetition, the total number of variational parameters required for this ansatz is
\begin{equation*}
  N_P = N_q \times (N_{\text{rep}} + 1) = N_s \times (N_b + 1) \times (N_{\text{rep}} + 1).
\end{equation*}
For the unreasonably small number of qubits shown in \fig{fig:ansatz}, $N_q = 4$, eight parameters are needed for the first repetition, with each subsequent repetition adding another four.

\begin{table}
  \begin{subtable}[t]{0.32\linewidth}
    \centering
    \caption{\label{table:vqe_a}Linear prepotential}
    \begin{tabular}{c @{\hspace{1.0\tabcolsep}} c @{\hspace{1.0\tabcolsep}} r @{\hspace{1.0\tabcolsep}} r}
      \hline
      N & \La &  Exact~~~ &  VQE~~~   \\
      \hline\hline
        &   2 &  6.97e-03 &  6.97e-03 \\
        &   4 &  3.24e-05 &  6.61e-05 \\
      2 &   8 &  6.89e-07 &  1.08e-01 \\
        &  16 & -2.54e-06 & ---       \\
        &  32 & -6.28e-06 & ---       \\
      \hline
        &   2 & -9.97e-02 & -1.28e+00 \\
      3 &   4 &  1.21e-04 &  4.99e-01 \\
        &   8 &  1.46e-05 & ---       \\
      \hline
      4 &   2 & -2.08e-01 & ---       \\
        &   4 & -6.07e-05 & ---       \\
      \hline
    \end{tabular}
  \end{subtable}\hfill
  \begin{subtable}[t]{0.32\linewidth}
    \centering
    \caption{Quadratic prepotential, $c=-0.2$}
    \begin{tabular}{r @{\hspace{1.0\tabcolsep}} r}
      \hline
       Exact~~~ &  VQE~~~   \\
      \hline\hline
      -9.11e-01 & -9.11e-01 \\
       1.82e-01 &  2.26e-01 \\
       1.31e-01 &  7.49e-01 \\
       1.95e-01 & ---       \\
       1.96e-01 & ---       \\
      \hline
      -1.28e+00 & -1.28e+00 \\
       3.02e-01 &  5.08e-01 \\
       2.70e-01 & ---       \\
      \hline
      -1.77e+00 & ---       \\
       3.27e-01 & ---       \\
      \hline
    \end{tabular}
  \end{subtable}\hfill
  \begin{subtable}[t]{0.32\linewidth}
    \centering
    \caption{Quadratic prepotential, $c=-0.8$}
    \begin{tabular}{r @{\hspace{1.0\tabcolsep}} r}
      \hline
       Exact~~~ &  VQE~~~   \\
      \hline\hline
      -9.11e-01 & -9.11e-01 \\
      -1.15e+00 & -1.15e+00 \\
      -1.73e-02 &  6.89e-01 \\
       1.95e-02 & ---       \\
       1.94e-02 & ---       \\
      \hline
      -1.28e+00 & -1.28e+00 \\
      -1.15e+00 & -1.10e+00 \\
      -1.89e-02 & ---       \\
      \hline
      -1.77e+00 & ---       \\
      -1.15e+00 & ---       \\
      \hline
    \end{tabular}
  \end{subtable}
  \caption{\label{table:VQE}Ground state energies from classical diagonalization and the VQE for the linear and quadratic prepotentials discussed in the text, with $N_s$ lattice sites and Dirichlet BCs.  The linear prepotential (\eq{eq:linear}) is expected to preserve supersymmetry with a vanishing ground state energy upon removing the cutoff, $\La \to \infty$.  For the quadratic prepotential (\eq{eq:quad}), supersymmetry is expected to break dynamically for $c > c_0 \approx -0.5$.}
\end{table}

Now that we have summarized the integredients used in our VQE computations, we can consider some results shown in Table~\ref{table:VQE}.
Here we record the minimum energy obtained across a modest number of independent VQE runs (at least 100) and compare this to the result of classical diagonalization.
For each of the linear prepotential (\eq{eq:linear}) and the quadratic prepotential (\eq{eq:quad}) with either $c > -0.5$ or $c < -0.5$, we consider Wess--Zumino systems with $N_s = 2$, $3$ and $4$ sites with Dirichlet BCs, and (depending on $N_s$) cutoffs up to $\La = 32$.
As a reminder, the ground-state energy should go to zero for sub-tables (a) and (c), where supersymmetry is expected to be preserved, while remaining non-zero for sub-table (b) where supersymmetry is expected to break dynamically.

From the classical computations in Table~\ref{table:VQE} we can see results consistent with these expectations, while the limitations of working with such small $N_s$ and \La are also clear (especially in the form of the negative energies allowed by finite $\La$).
In particular, we see that $\La \gtrsim 8$ is typically required in order for the ground-state energy for the quadratic prepotential with $c = -0.8$ to decrease below the non-zero constant approached due to spontaneous supersymmetry breaking for $c = -0.2$.
In part due to the Dirichlet BCs, both larger cutoffs and larger numbers of sites are needed for the ground-state energy to reliably approach zero for the free theory (linear prepotential).

More relevant for the current discussion are the VQE ground-state energy results in the table, which are unchanged compared to Table~1 in \refcite{Culver:2023iif}.\footnote{Table~\ref{table:VQE} corrects errors in the classical computations reported by \refcite{Culver:2023iif}, which used an iterative eigensolver and in some cases missed the ground state due to computing too few eigenvalues.  Here we have switched to a full (`dense-matrix') diagonalization of the hamiltonian.}
Even in the simplest case of the free theory, the VQE can struggle to reach the small $E_0$ predicted classically as the number of variational parameters increases.
In less-trivial cases, this could make it difficult to determine whether or not the ground state energy is really approaching zero --- that is, whether or not supersymmetry breaks spontaneously.

\subsection{Variational quantum deflation} 
This issue can be elegantly addressed by moving from the VQE to the VQD algorithm.
In essence, the VQD sequentially carries out some number of VQE solves, each time deflating the previous eigenstates.
That is, for the $k$th solve the objective function minimized by the VQD becomes
\begin{equation*}
  H_k = H + \sum_{i = 0}^{k - 1} \be_i \ket{\Psi_i} \!\bra{\Psi_i},
\end{equation*}
with the constraint that $\be_i > E_k - E_i$~\cite{Higgott:2019vqd}.
For simplicity we choose a constant $\be = 2$, which satisfies this inequality for the Wess--Zumino model systems we have analyzed so far.

\begin{figure}
  \includegraphics[width=0.48\linewidth]{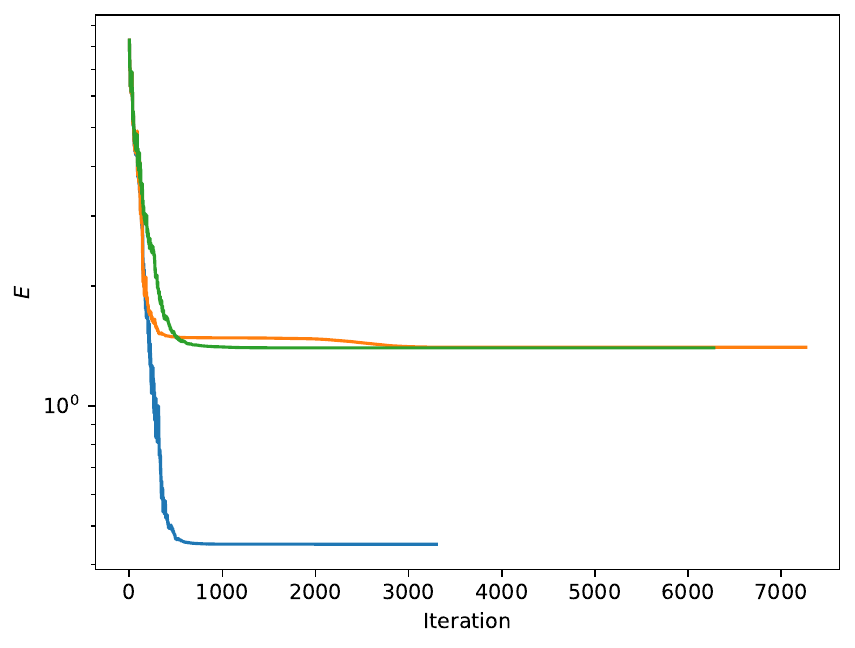}\hfill \includegraphics[width=0.48\linewidth]{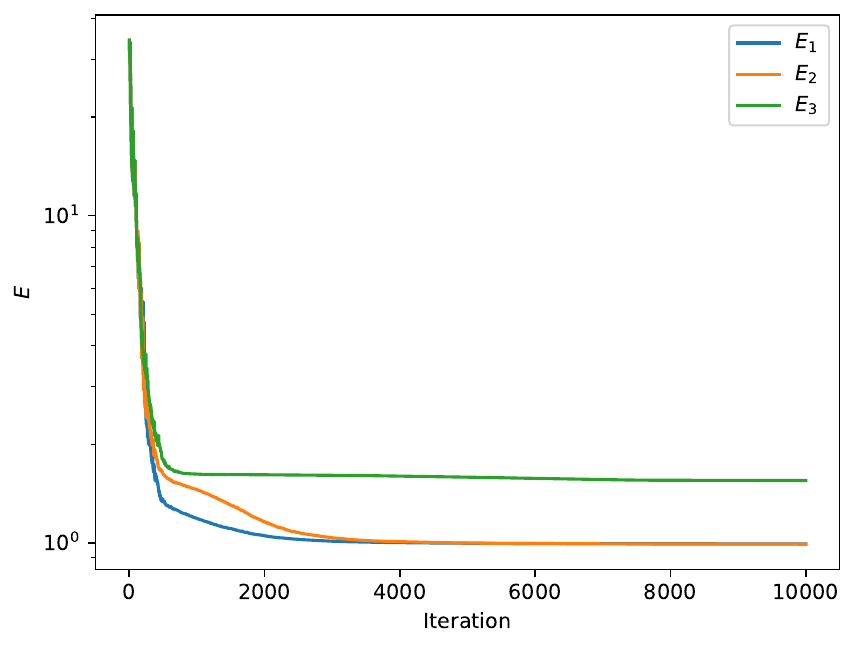}
  \caption{\label{fig:VQD}Results from representative VQD runs for the $\cN = 1$ Wess--Zumino model, estimating the three lowest energies with periodic BCs.  \textbf{Left:} For the linear prepotential with $N_s = 4$ sites and $\La = 4$, the isolated ground state is consistent with the expected preservation of supersymmetry.  \textbf{Right:} For the quadratic prepotential with $c = 0$, $N_s = 3$ sites and $\La = 8$, the paired ground states imply dynamical supersymmetry breaking.}
\end{figure}

Because the VQD is based on VQE solves, it will encounter the same issue with potential non-convergence to the true minimum energy of the objective function.
However, thanks to the properties of supersymmetric theories discussed in \secref{sec:lattice}, we don't need to rely on the precise values of the energies obtained by the VQD.
Instead, we can focus on the pairing structure of those energies, classifying this behavior into three categories.
First, if there is a large gap between a single lowest energy and pairs of higher energies, we can conclude that supersymmetry is preserved, independent of the actual values of these energies.
The left panel of \fig{fig:VQD} illustrates this situation for the linear prepotential with $N_s = 4$ sites, periodic BCs and $\La = 4$.
This is one of the largest systems we can diagonalize classically; its quantum simulation would require $N_q = 12$ qubits and $N_P = 36$ variational parameters.
While the energies found by this particular VQD run, $(0.450, 1.399, 1.405)$, are far from the $(-0.001, 0.829, 0.829)$ obtained from classical diagonalization, the qualitative pairing structure is clear.

Second, if instead all energies come in pairs, with no single lowest-energy state, we can conclude that supersymmetry is spontaneously broken.
The right panel of \fig{fig:VQD} illustrates this situation for the quadratic prepotential with $c = 0$, $N_s = 3$ sites and $\La = 8$, which leads to the same $N_q = 12$ qubits and $N_P = 36$ variational parameters as the left panel.
Again, the energies found by this particular VQD run, $(0.991, 0.993, 1.551)$, are far from the $(0.416, 0.416, 1.089)$ obtained from classical diagonalization, but the clear pairing structure provides a robust result nonetheless.

The third possibility is that the VQD may not actually find pairs of energies --- in other words, it may fail.
Had each VQE solve in the VQD run shown in the right panel of \fig{fig:VQD} been cut off after 1500 iterations, we might have ended up with this situation.
In this case, the run simply needs to be disregarded as inconsistent with the supersymmetric lattice hamiltonian.
Even here, the VQD provides an advantage over the VQE, where it would be more ambiguous whether or not a particular run had failed.
In practice, we find that most VQD runs result in clear pairs of energies, indicating that the explicit supersymmetry breaking introduced by the finite cutoff \La is not problematic.


\section{\label{sec:conc}Conclusions and next steps} 
The results of our explorations of quantum computing as a means to investigate dynamical supersymmetry breaking in the (1+1)-dimensional $\cN = 1$ Wess--Zumino model, while still preliminary, illustrate how clever algorithmic approaches may dramatically enhance the capabilities of existing and near-future quantum computers.
Specifically, by using the VQD to analyze the pairing structure of low-lying energies, rather than focusing on the ground-state energy itself obtained with the VQE, it becomes much easier to distinguish between systems with preserved vs.\ spontaneously broken supersymmetry.
This is a very encouraging recent development that we will exploit as we finalize our initial investigations.

There are of course many further directions we are eager to explore.
For example, as mentioned in \secref{sec:qubit}, it may be very interesting to make use of continuous-variable quantum computing for the Wess--Zumino model~\cite{Jha:2023ecu}.
Specifically, we are interested in a `hybrid' approach in which we continue to use qubits for the fermions $\psi_n$ while exploiting truncated `qumodes' to more efficiently represent the bosons $\phi_n$.
This approach seems likely to significantly reduce resource requirements for quantum simulation, and may also reduce the number of parameters needed in variational algorithms.
Refs.~\cite{Crane:2024tlj, Araz:2024kkg} exploring such hybrid approaches appeared while this proceedings was under review.

As our investigations mature and our current classical simulations of quantum devices indicate the most promising approaches for us to focus on, we are also eager to scale up our work using actual quantum hardware.
This will be crucial to better understand the capabilities of existing devices and explore the power of various error mitigation techniques in this context.
In parallel, we are turning our attention to real-time evolution.
While real-time evolution using quantum computers is more demanding than the variational approaches we have focused on so far, it could open up many additional analyses, including two-particle scattering and the identification of the massless goldstino expected to accompany dynamical supersymmetry breaking.
As our research advances, we can also look forward to generalizing beyond the Wess--Zumino model to more complicated (1+1)-dimensional supersymmetric field theories, including super-Yang--Mills~\cite{Catterall:2017xox}, super-QCD~\cite{Catterall:2015tta} and the supersymmetric Gross--Neveu--Yukawa model~\cite{Fitzpatrick:2019cif}.


\vspace{20 pt} 
\noindent \textsc{Acknowledgments:}~We thank Raghav Jha for discussions of qumodes, and Emanuele Mendicelli for many helpful conversations --- especially for pointing out errors in Table~1 of \refcite{Culver:2023iif}.
This work was supported by UK Research and Innovation Future Leader Fellowship {MR/S015418/1} \& {MR/X015157/1} and STFC grants {ST/T000988/1} \& {ST/X000699/1}. \\[8 pt]

\noindent \textbf{Data Availability Statement:} The raw data used in this work can be obtained by contacting DS.

\bibliographystyle{JHEP}
\bibliography{lattice23}
\end{document}